\documentstyle[12pt]{article}

\def\be{\begin{eqnarray}}
\def\ee{\end{eqnarray}}
\def\lb{\label}  
\begin{document}
\hfill {\tt hep-ph/0201248}
\vskip0.3truecm
\begin{center}
\vskip 2truecm
{\Large\bf
Classical and quantum cross-section for 
black hole
 \vskip0.5truecm
production in particle collisions
}\\ 
\vskip 1truecm
{\large\bf Sergey N.~Solodukhin\footnote{
e-mail: {\tt soloduk@theorie.physik.uni-muenchen.de}}
}\\
\vskip 0.8truecm
{\it 
Theoretische Physik,
Ludwig-Maximilians Universit\"{a}t, 
\vskip 0.2truecm
Theresienstrasse 37,
D-80333, M\"{u}nchen, Germany}
\vskip 1truemm
\end{center}
\vskip 1truecm
\begin{abstract}
\noindent
We suggest a simple model to study the problem of the black hole production
in particle collisions. The cross-section for the classical and
quantum production  is analysed within this model.
In particular, the possibility to form a black hole in collision of  low energy particles
(or at large impact parameter) 
via the quantum tunneling mechanism is pointed out.
It is found that, in this model, 
the geometric cross-section gives a good estimate for the production
at low and high energies. We also reconsider the arguments in favor of exponential
suppression for the  production of trans-Planckian black hole
and  conclude that no such suppression in fact appears.
Analyzing the probability for the black hole production  
we point out on the importance of the 
back-reaction and  reevaluate the contribution of the black hole formed in gravitational
collapse to the Euclidean path integral.
\end{abstract}
\vskip 1cm
\newpage
\section{Introduction}
\setcounter{equation}0

For long time black holes have been an interesting object for theoretical
exercises that lies on the intersection of General Relativity and Quantum Mechanics.
Many unusual features of this object have been revealed in the purely theoretical study 
in the last three decades. Most significantly, this includes the Hawking radiation and 
the thermodynamical description of the black hole \cite{BH}.

In spite of the considerable progress 
the fundamental theory of black holes never had a chance
to become a field of science having a touch with  experiment.
Nevertheless, 
the numerous attempts  to analyse the possibility to create a black hole
in laboratory have been made in the past \cite{tH}-\cite{Fabbrichesi:1993kd}.
It became clear that the problem of producing  a black hole in, say,
a particle collider is not just a problem of reaching high energy.
The analysis of collisions at trans-Planckian energy in the so-called eikonal regime
reveals no black hole production. In this regime the particles fall to each other at
large transverse separation and do not produce strong gravitational field
in the transverse direction.
The problem thus is in overlapping 
the wave functions of the colliding
particles close enough to turn on the strong gravitational interaction. 

On the other hand, the total effect is usually expected to be visible only when
the energy of the  particles becomes Planckian or even trans-Planckian. 
Since the Planckian scale lies far beyond the Standard Model scale, 
the problem of producing black holes in particle collisions was always considered 
of a purely
academic interest. This situation has a chance to change. There were  suggested 
phenomenologically viable models  \cite{Arkani-Hamed:1998rs},\cite{Antoniadis:1998ig} 
of multi-dimensional gravity
in which the Planck scale can be as low as in a TeV range. This opens an exciting
(yet requiring further examination) possibility to enter the Planckian
region in realistic future colliders and makes the black hole production
one of the most urgent theoretical problems. That is why the issue of 
producing black holes in LHC or by cosmic rays becomes increasingly popular 
\cite{Giddings:2001bu}-\cite{Kowalski:2002gb}.

As for the theoretical  aspects  of this problem they are 
surprisingly undeveloped. Estimating  the cross-section 
for the production of black hole 
in collision with total energy $E$
one typically takes it to be given by the horizon area \cite{Banks:1999gd}, 
$\sigma_{\tt prod}\sim \pi r^2_g$, 
where $r_g=2GE$ is the radius of the created black hole. This is the so-called geometric
cross-section. 
The derivation of this estimate is based mainly on intuition and was not actually 
justified theoretically. It is also a part of general believe that the gravitational collapse
at trans-Planckian energies does not much differ from the collapse as we know it in
General Relativity. Indeed, the expected radius $r_g\sim GE$ of the produced black hole
is much bigger than the Planck length. Therefore, the Quantum Gravity corrections
can be considered as small and not playing the key role in this process.

This, however, does not guarantee that the naive picture is automatically correct.
The validity of the geometric cross-section was criticized in \cite{Voloshin:2001vs} (see also 
\cite{Voloshin:2001fe}) as not actually providing us with the correct value for the semiclassical
cross-section.  It was argued in \cite{Voloshin:2001vs} that semiclassical treatment
of the problem reveals the exponential suppression by the factor $\exp (-4\pi G E^2 )$.
If correct, such suppression would make negligible the production of trans-Planckian 
black holes. This conclusion contradicts the intuition based on the classical 
collapse picture. 
It is therefore principally important to better understand this issue.

In this paper we partially fill the gap in the theoretical study and give 
a classical and quantum analysis of the problem of
black hole production in particle collisions in a simple (still meaningful)
model. 
An advantage of this model is that it suggests a well-defined procedure for actually computing
the classical and quantum cross-section for the black hole production.
In particular, we obtain a justification for
the geometric cross-section in a wide range of the energy.
For lower energy particles our model suggests an interesting possibility
to produce black hole in a purely quantum process when particles tunnel through the 
effective potential
barrier separating them in the radial direction. Since the  wavelength
of such particles is much bigger than the critical impact parameter a quantum treatment
(like the one we perform) is necessary. The cross-section for this process is 
given by  the geometric cross-section.
 
We also analyse in detail the arguments of paper \cite{Voloshin:2001vs}
in favor of the exponential suppression.
Two approaches were suggested in \cite{Voloshin:2001vs}: using the path integral
and the statistical approach. We reconsider both approaches and 
find no such suppression to actually appear. Our counter-arguments are quite instructive
and include the important issues of backreaction of the Hawking radiation 
as well as correct evaluation of the 
contribution of the black hole formed in gravitational collapse to the Euclidean
path integral.

\section{A model picture for the black hole production}
\setcounter{equation}0

The difficulties of the analysis of the black hole formation in the process
of colliding two particles lies in the fact that the classical 
two-body problem is not solved in General Relativity (the only exception is
$(2+1)$-dimensional case \cite{Matschull:1998rv}, \cite{Louko:1999hk} ; 
it may serve to guide our intuition
in higher-dimensional case but should be taken with some caution: gravity in 
3 dimensions does not propagate and thus describes a very special 
type of interaction). In order to overcome this difficulty  we 
suggest a simple model in which the black hole production 
(or, actually, a process equivalent to it) can be studied in a rather straightforward
way.

Let us start with a classical picture of 
two particles  with energy $E_1$ and $E_2$ respectively  moving towards
each other at certain impact parameter $b$.  For simplicity we
will be considering ultra-relativistic particles ($m_i/E_i<<1$)
with velocity close to the speed of light. Also, we assume that
the motion of the system as whole happens in one plane.
In Newtonian mechanics this system would be equivalently described 
in the center mass coordinate system  as a test particle with 
energy $\omega={E_1 E_2\over E_1+E_2}$
falling on the gravitating center of mass $M=E_1+E_2$. In General Relativity 
this picture is more complex due to the non-linear nature of the gravitational 
interaction. In order to include the gravitational interaction in the game
and put everything into a still analytically tractable scheme we
replace the above picture with an approximate one. We consider a gravitating center
with mass $M$ which creates around it  the gravitational field described 
by Schwarzschild metric with gravitational radius $r_g=2GM$, and a test particle 
with energy $\omega$ falling into the center from infinity at 
the impact parameter $b$. 
It should be noted that in this picture the black hole horizon at $r=r_g$
is absolutely fictitious. However, we say that the actual horizon forms when 
the test particle crosses $r=r_g$.
The latter thus signals for the black hole production in the original picture. 
This model is of course
an approximation. Its  main advantage though is that it suggests a well defined procedure to
compute, both classically and quantum mechanically,  the cross-section
for the black hole production when two particles collide. Also, 
the formulas relevant to this picture are already available in the literature studying the
classical and quantum particle scattering by black hole, 
a review of the existing literature can be found in
\cite{Frolov:wf}. 
We just have to give them a new interpretation.
However, this model is not a substitution for the desirable
analysis of non-linear process of actual gravitational collapse 
by the colliding particles. The earlier works in this direction include
\cite{Yurtsever:vd}, \cite{D'Eath:hb}, \cite{D'Eath:hd}, \cite{D'Eath:qu}.
An interesting  recent work is \cite{Eardley:2002re}. 
Our model should be considered as complimentary
to such analysis.

The classical radial motion of the test particle is determined 
by the geodesic equation (we put $c=1$)
\be
({dr\over dt})^2=(1-{r_g\over r})^2 b^2\left({1\over b^2}-{1\over r^2}
(1-{r_g\over r})\right)~~,
\lb{2.1}
\ee
where we used the fact that for a ultra-relativistic
particle, its energy $\omega$ and angle momentum $L$ 
(computed with respect to the gravitating center)
are related as $L/\omega=b$. 
For  massive particle there is an extra term in the geodesic 
equation (\ref{2.1}).
However, in the  ultra-relativistic ($m/\omega<<1$) limit this term can be neglected if 
the impact parameter 
satisfies condition $b>>Gm$ thus excluding values of the angle momentum close to zero. 
Note that for the present situation   the black hole mass 
$M\sim \omega$ and hence $(Gm)/r_g<<1$. So that it does not put a serious
restriction to our model.
The geodesic trajectories described by (\ref{2.1})
are well studied. For a particle coming from infinity the crucial relation is
the relation between $1/b^2$ and the maximal value of the
effective potential $V(r)={1\over r^2}(1-{r_g\over r})$.
This potential takes its maximal value at $r_m={3\over 2}r_g$ and
$V(r_m)={4\over 27 r_g^2}$.  Therefore, for impact parameter
$b<b_{\tt cr}=3\sqrt{3}/2 r_g$ the test particle  is 
captured by the gravitating center and eventually falls into  horizon. 
Having in mind the original
picture of the colliding particles we  say that there forms a black hole with 
horizon radius $r_g$.  In the case $b>b_{\tt cr}$ the particle just scatters 
off the center and the gravitational capture does not happen.
In the original picture, this would correspond to particles passing each other 
without actually forming  the black hole.  Thus, classically, the cross-section 
for the black hole formation is given by
\be
\sigma_{\tt cl}=\pi b^2_{\tt cr}={27\over 4} \pi r^2_g=27\pi G^2 M^2~~.
\lb{2.2}
\ee
Thus, this model predicts that, classically, two colliding particles 
form a black hole if they pass each other at the shortest distance
$b<b_{\tt cr}=3\sqrt{3}/2 r_g$ (where $r_g$ is the gravitational radius for the
system of these two particles). Note, that this is a little bigger than
one could expect from, say, Thorne's hoop conjecture. 
This is because the particles need to get over  the
potential barrier separating them and staying just outside the 
effective horizon. Then, having reached  the other 
side of the barrier they fall to each other without any other obstacles.

This picture can  be also  analysed quantum mechanically.
The relevant processes, black hole scattering and absorption, were well studied
in the past and here we give a brief summary with the re-interpretation
according to our picture of the black hole production.
Decomposing quantum field in spherical harmonics, $\Phi_{lm}\sim {u_l(r,\omega )\over r}
e^{-i\omega t}Y_{lm}(\theta, \phi )$, one arrives at the radial wave equation
\be
\left({d^2 \over dr^2_*}+\omega^2-U_l(r)\right) u_l(r,\omega)=0,
\lb{2.3}
\ee
which is a quantum mechanical analog of the classical equation (\ref{2.1}).
We denote ${d\over dr_*}=(1-{r_g\over r}){d\over dr}$ and
\be
U_l(r)=(1-{r_g\over r})\left({l(l+1)\over r^2}+{r_g (1-s^2)\over r^3}\right)~~,
\lb{2.4}
\ee
where $s$ is spin of the particle (below we put $s=0$),
is the effective potential similar to the potential in eq.(\ref{2.1}).
The equation (\ref{2.3}) describes waves scattering by the potential (\ref{2.4}).
The relevant modes are defined to be in- and out-going at infinity $r_*\rightarrow \infty$,
$u_l(r,\omega )\sim A_{\tt out}(\omega )e^{i\omega r_*}+ A_{\tt in}
(\omega )e^{-i\omega r_*}$,
and  only out-going at horizon $u_l(r,\omega )\sim e^{-i\omega r_*}$, 
$r_*\rightarrow -\infty$. Again, in this picture we say that the black hole 
production takes place if the wave is absorbed by the effective horizon
around the gravitating center of mass $M$.

The probability of  wave to penetrate through the potential barrier is 
$$
\Gamma_{l,\omega}=1-|{A_{\tt out}\over A_{\tt in}}|^2~~.
$$
Note, that it is probability for a spherical wave. Since the actual wave at infinity 
looks more like a plane wave
the latter should be decomposed on the spherical modes
\be
&&e^{-i\omega z}=\sum_{l=0}^\infty K_{l}(\theta )Y_{l0}(\theta,\phi )~~, \nonumber \\
&&K_l(\omega )={i^l\over 2\omega} [4\pi (2l+1)]^{1/2}~~.
\lb{2.5}
\ee
The  cross-section to capture the test particle is then defined as follows
\be
\sigma_{\tt quant}=\sum_l |K_l|^2\Gamma_{l,\omega }~~.
\lb{2.6}
\ee
It is also convenient to consider each partial wave characterized by
angular momentum $l$ as falling on the gravitating center at impact
parameter $b$,
$$
b=(l+{1\over 2}){1\over \omega }~~.
$$
In the high energy limit the geometrical optics analysis based on
equation (\ref{2.1}) becomes a good approximation.
In this limit the cross-section (\ref{2.5})
approaches the classical value (\ref{2.2}). One can also compute
$1/\omega^2$ -corrections. The result then reads
\be
\sigma_{\tt quant}\simeq {27\over 4}\pi r^2_g-{2\over 3}\pi {1\over \omega^2}~~.
\lb{2.7}
\ee
The horizon absorbs partial waves with impact parameter $b\leq {3\over 2} r_g$
in this regime.

It is also interesting to analyse  the opposite limit of small $\omega$.
As argued in \cite{M} it is actually the limit of small $\omega r_g/l$,
i.e. $b>>r_g$ in this case. Remarkably, the cross section (\ref{2.6})
approaches a finite number in this regime equal to the  horizon
area\footnote{This is a quite old result first obtained by Starobinsky \cite{Starobinsky}, 
see also a nice analysis given in a paper by Unruh \cite{Unruh:fm}. 
The higher-dimensional analysis is given recently in
\cite{Das:1996we}.}
\be
\sigma_{\tt quant}\simeq \pi r^2_g~~.
\lb{2.8}
\ee
Naively, one would  expect this to vanish.  
The transmission probability $\Gamma_{l,\omega }$ does vanish for small $\omega$
as $\sim \omega^{2l+2}$ and dominates for s-wave. However, the contribution
of the s-wave in the in-falling plane wave diverges  $|K_l|^2\sim {1\over \omega^2}$,
as is seen from (\ref{2.5}). 
The two tendencies compensate each other in (\ref{2.6})
for the s-wave. This results in the finite cross section (\ref{2.8}). 
Since only the wave that approaches the horizon radially 
is absorbed  it is not surprising that the cross-section becomes equal to the
horizon area.
We should
also note that the penetration through the barrier of the wave 
with small $\omega$ (or large impact parameter, $b>>r_g$) is classically forbidden!
Therefore, the cross section (\ref{2.8}) is entirely due to
the quantum tunneling effect.

We see that the geometric cross-section gives a rather good estimate
for the black hole production in  a wide range of energy $\omega$.
The ratio of the actual cross-section and the geometric cross-section appears to be
a slow changing function of  the energy.

Does this simplified picture give a correct description of what happens in the
actual gravitational collapse when two particles collide?
We do expect this picture to be a good approximation when one of the particles has much bigger 
energy than 
another. The cross-section in this case is given by (\ref{2.8}) and can be quite large if
one of the particles has trans-Planckian energy.

In the case of the approximately equal energy this picture is still valid
when two particles collide at the impact parameter of the same order as the effective
gravitational radius. The gravitational field is strong and essential
when particles come close to each other at distance $r\sim r_g$ and the total gravitational field
indeed can  be approximated by the field created by total mass located somewhere in between.
This is also the case for the high-energy collision. In fact, in this regime
the geometrical optics analysis becomes a good approximation. Therefore,
the validity of our model then reduces to the same question for the classical
problem of two colliding bodies in General Relativity. In case of colliding black 
holes this issue was studied in the literature (see for instance \cite{Anninos:1995vf} 
and references therein) 
and good agreement with exact (numerical) solution was found.

The validity of the picture is less obvious when the impact parameter is relatively large
compared to 
the gravitational radius $r_g$. Still, the effective potential we considered 
models the actual potential between
the particles. Classically, the collapse is forbidden
in this case.
However, we think that our analysis opens an important
possibility to produce a black hole by colliding low energy particles
via the quantum tunneling. 
This regime is closer to the real situation when the wavelength of the particles is much bigger 
than the critical
impact parameter thus justifying the quantum mechanical treatment we performed.

There are few directions for further improving our calculation. First of all, the black hole
formed in collision of particles is expected to be rotating. Therefore, 
the rotation should be included in our picture. The corresponding classical and quantum
analysis is given in the existing literature, see for example \cite{Dym} and
\cite{Frolov:wf},  the extension of our consideration 
to this case is quite straightforward. 

The Quantum Gravity corrections neglected in the above consideration can be taken into account.
These corrections result in shifting the location of the Schwarzschild horizon, so that the new location 
$r_{\tt q}$ is given by (an example of the quantum corrected Schwarzschild metric is given in \cite{Kazakov:1993ha})
\be
r^2_{\tt q}=(2GM)^2+\#~ l^2_{\tt pl}~~,
\lb{*}
\ee
where $l_{pl}\sim\sqrt{G}$ is Planck length and $\#$ is some number which may also include
the slow changing with energy logarithmic  term $\ln M$.  The geometric area is now given by $\pi r^2_{\tt q}$.
If the total energy $M$ is much bigger than $M_{\tt pl}\sim G^{-{1\over 2}}$ the quantum shift
can be neglected. However when $M\simeq {\tt few}~ M_{\tt pl}$ (it is expected that the minimum mass to produce 
black holes on LHC is $M_{\tt min}\simeq 5 M_{\tt pl}$, see \cite{Giddings:2001ih})
 both terms in (\ref{*}) become equally important. This may change the numerical
value of the cross-section (see also discussion in \cite{Giudice:2001ce}).

Another important issue missed so far is the gravitational radiation.
Indeed, the test particle falling onto the black hole is expected to radiate
some amount of its energy via producing the gravitational waves
(gravitons). The standard estimate for this amount is
\be
E_{\tt rad}\sim \kappa ~ \omega^2/M~~,
\lb{2.9}
\ee
where $\kappa\simeq 0.01$ in non-relativistic case (see \cite{Anninos:1995vf});
for ultra-relativistic particles (or black holes) of equal mass one has
$\kappa\simeq 0.6$ and $\kappa\simeq 0.2$ if $\omega$ is much less than $M$ 
(see \cite{Smarr}). The  energy loss due to radiation
in the head-on ultra-relativistic collision
can be as large as $25\%$ of the total energy (\cite{D'Eath:hb}).
The main part of this energy goes to the quasi-normal oscillations.
The latter are excited as the particle passes through the peak
of the potential barrier. 
For high energy this peak is at $r={3\over 2}r_g$.
The actual mass of the
black hole formed by colliding the particles can be a certain fraction of
the value $M=E_1+E_2$ which results in a smaller value for $r_g$.
So that our formulas should be corrected respectively due to this energy loss.

Interestingly, in the case of low energy collision (i.e. regime in which
$\omega <<M$)  when  
the particles tunnel through the barrier separating them they do not actually pass
through the peak of the potential barrier and hence lose less energy.
This is also seen from (\ref{2.9}). 
Therefore, in some cases it might be more efficient to produce  black holes
by lower energy particles
via the quantum tunneling rather than in the high energy collapse. Note again that
the cross-section (\ref{2.8}) can be quite large in this regime.

\section{Is the black hole production  exponentially suppressed?}
\setcounter{equation}0

The applicability of the geometric cross-section to the process of the black hole production
in collision of particles with energy exceeding the Planck energy
was criticized in \cite{Voloshin:2001vs}, \cite{Voloshin:2001fe}.
It was argued in these papers that the semiclassical analysis gives rise to the
exponential suppression of the cross-section as
\be
\sigma_{\tt semicl}\sim e^{-4\pi G M^2}
\lb{3.1}
\ee
that makes negligible the production of black hole with large mass $M>>M_{\tt pl}$.
Note that this conclusion is quite counter-intuitive. 
Indeed, our experience based on the study of classical equations  
of General Relativity says that the concentration of large energy in small enough space-time region
inevitably leads to formation of a black hole. Also, the fact that the ratio
of $M/M_{\tt pl}$ is large means that the expected size $r_g$ of horizon  is much bigger than
the Planck length $l_{\tt pl}\sim \sqrt{G}$ 
and hence the Quantum Gravity corrections to the classical
process must be small not producing the exponential factors as in (\ref{3.1}).
The latter argument also means that the semiclassical
analysis  should be a reliable approximation to describe this process.
Note, that for the model considered in previous section the validity of the semiclassical
approximation for high energy justifies the fact that
the classical geometric optics analysis (\ref{2.1}), (\ref{2.2}) gives the
right answer in  this case. No exponential factor like (\ref{3.1}) arises there.

In this section we re-consider the arguments of paper \cite{Voloshin:2001vs}
and show that the consistent treatment of the problem does not lead
to the appearance of exponential factors similar to  (\ref{3.1}).
Two approaches were suggested in \cite{Voloshin:2001vs}.

\subsection{Statistical approach}

The black hole of mass $M$ can be viewed as a macroscopic 
object realized by a large number of micro-states ($H$),
${\cal N}=\exp (S_H)$, determined by the entropy $S_H=4\pi G M^2$.
The probability to create the black hole in collision of few particles
can be obtained by summing up the probabilities to create a black hole at a given 
micro-state $H$. Thus, the total probability is proportional to $\cal N$. 
On the other hand, for a given $H$ each such probability, by the CPT symmetry, is related to
the probability of the reverse process of the black hole  decay
into few (anti)particles. The latter can be estimated by the Gibbs formula
provided the black hole decays thermally with temperature $T_H=1/(8\pi GM)$.
These reasonings led in \cite{Voloshin:2001vs} to derive  the total probability in the form
\begin{eqnarray}
&&P({\tt few \rightarrow black~hole}) \sim {\cal N}~P({\tt  black~hole \rightarrow few}) 
\nonumber \\
&&\sim \exp {(S_H-\sum_i{E_i\over T_H})}~~,
\lb{3.2}
\end{eqnarray}
where $E_i$ are the energies of individual particles the black hole decays to.
The first term in (\ref{3.2}) is due to the black hole degeneracy
while the second term is the  probability of the reverse (decay) process\footnote{In
\cite{Giddings:2001ih} it was argued that the time reversal process should involve a white hole 
rather than a black hole. We leave aside this possibility assuming that what should stay in
(\ref{3.2}) is indeed the decay probability 
of the object formed in the direct process, i.e. of the black hole.
See also discussion in \cite{Voloshin:2001fe}.}.
The black hole decays until it disappears, therefore $\sum_i E_i=M$.
Then, what stands under the  exponent in (\ref{3.2}) is
$(S_H-T^{-1}_H M)$ which is equal to $(-S_H)$ for Schwarzschild black hole.
This is what was obtained in \cite{Voloshin:2001vs}.

What is missing in the above consideration is the fact that a radiated particle 
causes certain effect of the back-reaction  on the black hole. Namely, mass of the black hole
and its temperature change respectively. This effect is not difficult to take into account.
By the moment the black hole of initial mass $M$ has radiated particles with total energy $\omega$
its mass becomes ($M-\omega )$ and the inverse Hawking temperature is
$T^{-1}(\omega )=8\pi G (M-\omega )$. Therefore, the probability to radiate
next particle with small energy $d\omega$ is
$$
P( \omega, d\omega )\sim \exp (-T^{-1}(\omega )d\omega )~~.
$$
The total decay probability then is the product of these probabilities for all individual  
particles until the black hole disappears completely. The equation (\ref{3.2}) thus should be
replaced by the following equation
\be
P({\tt few \rightarrow black~ hole}) \sim \exp (S(M)-\int_0^M 8\pi G (M-\omega )d\omega )~~.
\lb{3.3}
\ee
It is easy to check that the expression under the exponent vanishes identically for the
Schwarzschild black hole. 

One might argue \cite{V3} that in our modification of the calculation given in \cite{Voloshin:2001vs}
the decay of black hole is a slow step-wise decay while 
for the process of the formation of black hole in 
few-particle collision  it is more relevant to consider 
a decay into   few (e.g. two) particles at once. 
However, it is clear that the latter  type of decay   can be  hardly 
called thermal and hence 
it is not eligible  at all to use (as it was done in (\ref{3.2})) the thermal form
for the probabilities. Moreover, after a minor modification
our calculation can be applied to a decay on arbitrary large pieces not assuming at 
all the thermal character of the decay. 
Indeed, on general grounds,
for a highly degenerate system the probability to radiate a particle
is proportional to the exponent of the corresponding change of the 
entropy\footnote{Note, that a similar idea has been used in  \cite{Parikh:1999mf}
to describe the Hawking radiation when the back-reaction of the radiated particles 
is taken into account.}. Applying this to black hole, we find that after 
the $i$-th particle has been radiated the entropy of black hole  changes on $-\delta S_i$
so that the discussed probability  is
\be
P({\tt few \rightarrow black~ hole}) \sim \exp (S_H-\sum_i \delta S_i )~~,
\lb{3.4}
\ee
where number of radiated particles can be arbitrary (two, three or a hundred).
Since the black hole is supposed to radiate completely in the reverse process, 
we have that $\sum_i \delta S_i=S_H$.
This again means no exponential suppression.

We should note that it was actually  expected in \cite{Voloshin:2001vs} that the 
back-reaction may
play certain role in the discussed semiclassical calculation 
and lead to some modification of the formula for probability.
However, this was argued to make a  relatively small effect which just changes the 
possible pre-factor in (\ref{3.1}).
We see however that the back-reaction is in fact crucial for the considered process
and removes the  exponential factor completely.

\subsection{Path integral approach}

In this approach, computing the probability (or transition amplitude) for the 
desirable process ($\tt few~$ $\tt particles \rightarrow black~hole$)
it is suggested in \cite{Voloshin:2001vs} that one has to deal with the path integral
over metric and matter fields subject to appropriate conditions
at $t=-\infty$ (few colliding particles) and at $t=+\infty$ (black hole).
Evaluating  path integrals one normally shifts the $t$-integration to the complex
half-plane: $t\rightarrow t-i\tau$. This results in considering the Euclidean 
section of the space-time. In the analysis given in  \cite{Voloshin:2001vs}
it is assumed that
the Euclidean section representing the contribution of the black hole is 
the eternal black hole instanton and the path integral  is given semiclassically
by
\be
P({\tt few \rightarrow black~ hole}) \sim \exp (-I_E[g])~~,
\lb{3.5}
\ee
where $I_E[g]$ is the gravitational action evaluated on the instanton.
The Euclidean black hole instanton can be viewed as a section of complex 
space-time of eternal black hole by the plane $t=0$ passing through the
bifurcation point at $r=r_g$. The instanton then is known to be regular manifold
with abelian isometry generated by vector $\partial_\tau$. This isometry has 
a stationary point at $r=r_g$. The regularity at this point requires the Euclidean time
$\tau$ to be periodic with the period being $8\pi GM$ for the Schwarzschild 
black hole. The Euclidean action then reduces to the boundary term 
\be
I_E[g]=-{1\over 8\pi G}\int_{r=\infty} (K-K_0)~~
\lb{3.6}
\ee
evaluated over boundary at $r={\tt const}
\rightarrow \infty$, $K$ is the extrinsic curvature
of the boundary. In order to regularise the gravitational action one normally subtracts 
the flat space contribution $K_0$.
Defined in this way action (\ref{3.6}) is known \cite{Gibbons:1976ue} to be equal to 
$$
I_E[g]=S_H=4\pi GM^2~~.
$$
This again seems to indicate the 
exponential suppression (\ref{3.1}) of the semiclassical probability.
According to \cite{Voloshin:2001vs} this is due to the exponentially small contribution 
of the black hole to the total probability.

The above analysis is perfectly suitable to describe the
spontaneous creation of a black hole (or, say, creation of 
black hole pairs in external field, see for example \cite{Dowker:up}, \cite{Hawking:1994ii})
when the created space-time can be matched to (a part of)
the Penrose diagram
of eternal black hole including the bifurcation point.
The exponential semiclassical 
suppression then is  expected since the process is forbidden 
classically. This is however not the case for the black hole 
formed in gravitational collapse\footnote{Otherwise, it would be 
applicable to the gravitational collapse of stars making the formation of black hole 
in this process quantum mechanically impossible.}.
In this case the resultant space-time which we could call the black hole
matches only to certain region of the Penrose diagram of the eternal black hole
that does not include the bifurcation point. 
Therefore, the Euclidean instanton of eternal black hole is not appropriate for this situation.
Considering a
$t={\tt const}$ slice of the total space-time
and extending it to a slice of a complex space-time
we find that it covers only a part of the
Euclidean instanton for $r\geq r_g+\epsilon$, where $\epsilon$ is non-zero
quantity measuring on $t={\tt const}$ slice the distance between the ``surface'' of 
the collapsing system and the  would-be  horizon. For certain value of $t$ this distance
vanishes and the black hole actually forms.
In the case under question the ``surface'' would be formed by trajectories of
the colliding particles. The region $0\leq r\leq r_g+\epsilon$ is inside the
collapsing ``body''. It can be modeled by  space with flat metric.
The complex space-time and its Euclidean section have the usual meaning for flat space.
The Euclidean instanton relevant to the collapse
is thus a part of the eternal black hole instanton 
and a flat disk glued together at $r=r_g+\epsilon$.
The extrinsic curvature has a jump on the surface $r=r_g+\epsilon$.
This jump can be thought as arising  due to the stress tensor of the collapsing matter.
This is the picture arising in the collapse of  spherical
shell \cite{Werner}. 
Evaluating the gravitational action the jump in the extrinsic curvature should be taken
into account,
so that we arrive at the action
\be
I_E[g]=-{1\over 8\pi G}\left(\int_{r=r_g+\epsilon}(K-K_0)+\int_{r=\infty} (K-K_0)\right)~~,
\lb{3.7}
\ee
where all $K$ are defined with respect to the normal vector directed to large $r$.
In order to single out the contribution of the hole itself we  take the limit 
$\epsilon\rightarrow 0$. The integral of $K_0$  at $r=r_g+\epsilon$ vanishes while that of
$K$  is  non-zero and 
equals  to the minus entropy of the black hole \cite{Barvinsky:1995dp}
$$
-{1\over 8\pi G}\int_{r=r_g+\epsilon}K=-S_{H}~~.
$$
Since the contribution of the external boundary is the same as before,
$+S_H$, we conclude that the gravitational action (\ref{3.7}) vanishes identically
(a similar calculation in the context of the stretched horizon approach (membrane paradigm) 
was done in \cite{Parikh:1997ma}).
This removes the dangerous exponent in (\ref{3.5}) and makes no
suppression to the probability.
Clearly, this is because the actual contribution of the black hole (formed in the
gravitational collapse) to the total probability is of order of one.
It is also consistent with the fact that the considered process of the black hole production via
gravitational collapse (the collision of particles is an example of such process)
is classically allowable (see also \cite{Giddings:2001ih}).


\section{Conclusion}
\setcounter{equation}0

The production of black holes in collision of particles is an interesting and
technically difficult problem. Complete solution should involve the classical and
quantum analysis of non-linear gravitational interaction between the particles.
In this paper we make a  step towards this solution and 
suggest a simple model in which the interaction is modeled by
certain effective radial potential. Within this model we analyse 
the cross-section for the production.
It is found that the geometric cross-section gives a rather good estimate 
at low and high energies. At low energy the process of the black hole production 
is forbidden classically and goes via the quantum mechanical mechanism of
the under-barrier tunneling.

In the second part of the paper we resolve the issue of the exponential suppression
of the trans-Planckian production of black holes.  Our analysis shows that
the consistent treatment of the approaches suggested in \cite{Voloshin:2001vs}
reveals no exponential suppression. This conclusion is based on the following
observations: i) the Gibbons-Hawking calculation  used in \cite{Voloshin:2001vs}
to estimate the black hole contribution to the Euclidean path integral is {\it not}
appropriate if the black hole was formed in gravitational collapse; ii) calculating the
probability of the reverse process of black hole decay
the backreaction of the radiated particles on the black hole geometry should be properly
included. Provided these issues are properly taken into account there appears no
exponentially small terms in the probability 
to produce black hole at trans-Planckian energy.

\bigskip

{\large\bf Acknowledgments}
At the initial stage of this work conversations with G. Dvali, 
G. Gabadadze, A. Vainshtein
and especially M. Voloshin were extremely useful. 
The later email correspondence with M. Voloshin is 
greatly acknowledged. I would like to thank M. Parikh for reading
the manuscript and his useful remarks. I thank the Aspen Center for Physics
for hospitality during the initial stages of this project.
This work would not be possible without the (always very generous) encouragement 
of my daughter Christina. This work is also  supported by the  
grant DFG-SPP 1096.

\end{document}